\documentclass[aip,apl,reprint,letterpaper,amssymb,amsmath,amsfonts,superscriptaddress,showkeys]{revtex4-1}
\usepackage[utf8]{inputenc}
\usepackage[T1]{fontenc}
\usepackage{lmodern}
\usepackage[american]{babel}
\usepackage{graphicx}
\usepackage{float}
\usepackage[activate={true,nocompatibility}]{microtype}
\usepackage[colorlinks=true, linkcolor=blue, citecolor=blue, urlcolor=blue]{hyperref}
\usepackage{eso-pic}

\begin{document}

\preprint{1}
\title{3D tomography of cells in micro-channels}

\author{S. Quint}
\email[]{Stephan.Quint@physik.uni-saarland.de}
\affiliation{Saarland University, Department of Experimental Physics, Campus E2.6 , 66123 Saarbrücken, Germany}

\author{A. F. Christ}
\affiliation{Saarland University, Department of Experimental Physics, Campus E2.6 , 66123 Saarbrücken, Germany}

\author{A. Guckenberger}
\affiliation{University of Bayreuth, Biofluid Simulation and Modeling, Department of Physics, Universitätsstraße 30, 95440 Bayreuth, Germany}

\author{S. Himbert}
\affiliation{Saarland University, Department of Experimental Physics, Campus E2.6 , 66123 Saarbrücken, Germany}

\author{L. Kaestner}
\affiliation{Saarland University, Department of Experimental Physics, Campus E2.6 , 66123 Saarbrücken, Germany}
\affiliation{Saarland University, Theoretical Medicine and Biosciences, Campus University Hospital, Building 61.4, 66421 Homburg, Germany}

\author{S. Gekle}
\affiliation{University of Bayreuth, Biofluid Simulation and Modeling, Department of Physics, Universitätsstraße 30, 95440 Bayreuth, Germany}

\author{C. Wagner}
\affiliation{Saarland University, Department of Experimental Physics, Campus E2.6 , 66123 Saarbrücken, Germany}


\date{\today}

\begin{abstract}
We combine confocal imaging, microfluidics and image analysis to record 3D-images of cells in flow.
This enables us to recover the full 3D representation of several hundred living cells per minute. Whereas 3D confocal imaging has thus far been limited to steady specimen, we overcome this restriction and present a method to access the 3D shape of moving objects. The key of our principle is a tilted arrangement of the micro-channel with respect to the focal plane of the microscope. This forces cells to traverse the focal plane in an inclined manner. As a consequence, individual layers of passing cells are recorded which can then be assembled to obtain the volumetric representation. The full 3D information allows for a detailed comparisons with theoretical and numerical predictions unfeasible with e.g.\ 2D imaging. Our technique is exemplified by studying flowing red blood cells in a micro-channel reflecting the conditions prevailing in the microvasculature. We observe two very different types of shapes: `croissants' and `slippers'. Additionally, we perform 3D numerical simulations of our experiment to confirm the observations. Since 3D confocal imaging of cells in flow has not yet been realized, we see high potential in the field of flow cytometry where cell classification thus far mostly relies on 1D scattering and fluorescence signals.
\end{abstract}

\maketitle

The optical detection and classification of biological cells is usually performed by two complementary approaches. On the one hand, optical microscopy is suitable for small sample amounts and steady specimen such as tissue slices.
Many techniques exist that allow to reveal textures at high resolutions on the sub-$\mu\text{m}$ scale.
For example, detailed 3D images of resting bio-particles such as red blood cells (RBCs) can be produced with confocal microscopes by scanning the $z$-axis of the objective in subsequent steps \cite{Khairy2008, Sakashita2012, Kim2014, Lanotte2016}. 
This technique also allows to highlight special structures by activating specifically dyed regions of interest.
However, even with fast spinning Nipkow disks, the record of moving objects such as cells in flow is highly challenging. 
Constraints mainly arise from the limited scanning performance of the mechanical actuators of microscopes. On the other hand, flow cytometry targets bio-particles suspended in flowing liquids and allows for high-throughput classification of objects \cite{Shapiro2003, Otto2015}.
Statistical information based on a few measurement parameters (fluorescence emission and scattering signals) finally serves to discriminate between different cell populations.

Obtaining the 3D shape of individual cells in flow has thus far been restricted to special circumstances such as when the cell performs a full rotation in the microscope's field of view \cite{Merola2017}. However, being able to analyze individual cells under general flow conditions is a very important task as it can give answers to basic questions concerning the physical properties of the cell membrane (elastic moduli, stress-free shape, etc.), or the preferred shape in various environments. Understanding the detailed behavior of individual cells also serves as a first step towards an in-depth comprehension of multi-particle interactions or even dense suspensions \cite{Forsyth2011, Kruger2013, Lanotte2016}.

\AddToShipoutPictureFG*{%
	\AtPageLowerLeft{% Start from the lower left corner of the page.
		\hspace{\dimexpr\oddsidemargin+1in\relax}% Move in the text by the size of the left margin: See comment at: http://tex.stackexchange.com/a/201500
		\raisebox{2\baselineskip}{% Move text up
			\parbox{\textwidth}{% Put the text into a parbox for line breaks.
				\footnotesize{}\textsf{% Formatting of the text
					EXAMPLE LINE: This is an author-created, un-copyedited version of an article accepted for publication in Applied Physics Letters. AIP Publishing is not responsible for any errors or omissions in this version of the manuscript or any version derived from it. The Version of Record is available online at \url{}.
}}}}}

\begin{figure}[t!]
	\centering
	\includegraphics{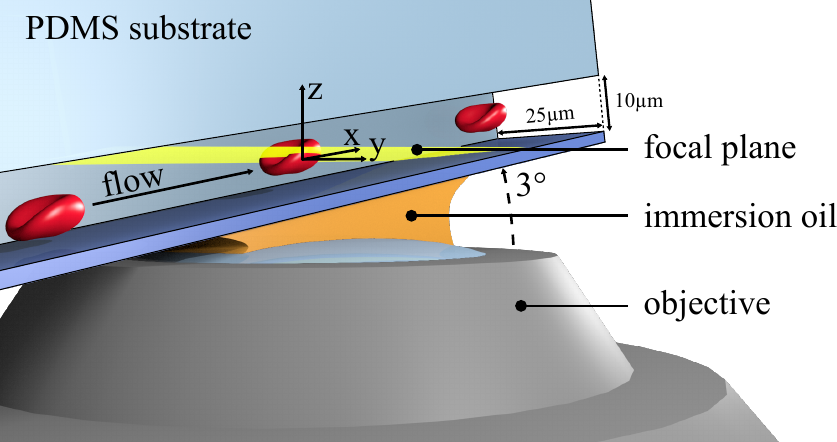}
	\caption{Sketch of the used setup. With a tilted stage, we incline our micro-channels with respect to the focal plane (yellow) of the objective such that we can take advantage of the objective's full field of view. Cells traverse the focal plane sectional-wise and a stack of image slices is recorded to recover the full 3D representation. Since movement of mechanical stages is not necessary, frame rates up to $600$ frames per second (FPS) can be realized. The used channel has a cross-section of $25\,\mu\text{m}\times{10}\,\mu\text{m}$ ($\text{width}\times\text{height}$) and is tilted by an angle of $\approx3^{\circ}$.}
	\label{fig1}
\end{figure}

\begin{figure*}[t!]
	\centering
	\includegraphics{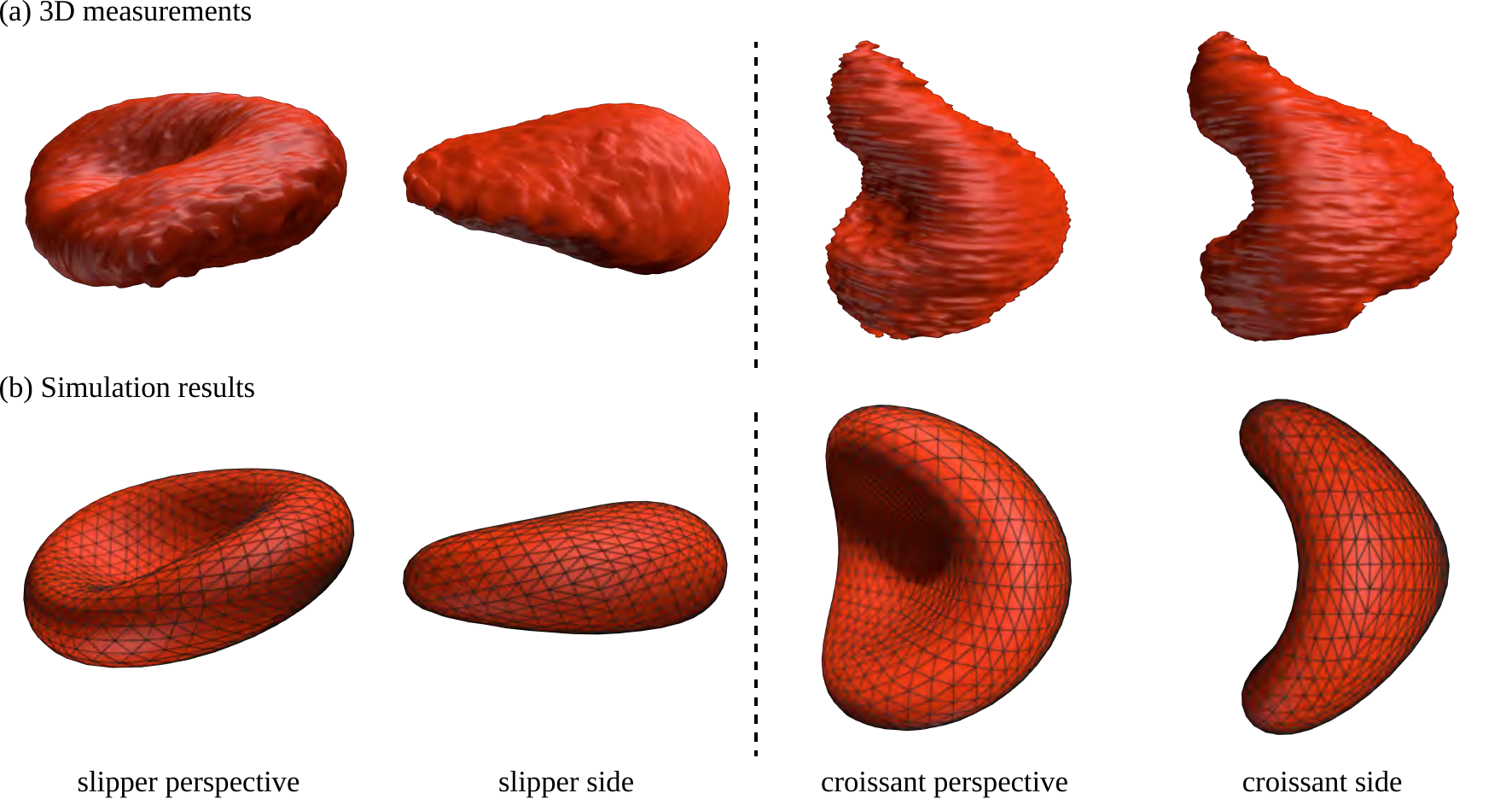}
	\caption{3D reconstruction of moving RBCs in micro-fluidic flow (top row). The channel dimension corresponds to $25\,\mu\text{m}\times10\,\mu\text{m}$ ($\text{width}\times\text{height}$). 3D assemblies are made up of about 250 cross-sectional images and show two possible RBC configurations at the same applied pressure drop: a slipper at $\approx 330\,\mu\text{m}/\text{s}$  and a croissant at $\approx 370\,\mu\text{m}/\text{s}$  (the difference in cell velocity results from a different vertical cell position). A perspective as well as a side view of the slipper and the croissant shapes are illustrated.  Pictures are rotated to give a proper impression on the cell geometry. However, since the rotation of cells relative to the flow are different, slices course from front to back in case of the slipper shape and from top to bottom for the croissant. Simulations of the respective cell shapes (bottom row) are generated using the periodic boundary integral method (see text). Giving a better impression on the volumetric representation, we refer the reader to our supplementary video material.}
	\label{fig2}
\end{figure*}

Here, we introduce a flow cytometry technique based on confocal imaging to record the \textit{three-dimensional} shape of \emph{flowing} cells in a micro-channel.
The basic idea is to not actuate the objective or sample stage, but rather to tilt the channel such that successive frames record different slices of the object moving through the focal plane of the confocal microscope (Fig.~\ref{fig1}). Depending on the cell velocity, about one hundred layers of a single cell can be recorded and then assembled to obtain a full 3D reconstruction. The acquisition speed is only limited by the rotation speed of the employed Nipkow disk and the frame rate of the used camera. Typically, a throughput of several hundred cells per minute can be handled and reconstructed by our technique. We exemplify our technique by considering the shapes of single human red blood cells (RBCs) in channel-sizes comparable to the structures found in the microvasculature. The attained shapes are of high importance for the macroscopic properties such as the pronounced shear thinning of blood.
Yet, current research on this topic was so far limited to 2D imaging methods \cite{Abkarian2008, Tomaiuolo2009, Coupier2012, Lanotte2014, Prado2015, Tomaiuolo2016, Claveria2016, Dupire2012, Basu2011} where the shapes are not always clearly identifiable \cite{Coupier2012}, 3D imaging of `frozen' cells where the method of freezing might influence the form \cite{Gaehtgens1980, Lanotte2016}, and 2D \cite{Secomb2007, Kaoui2009, Kaoui2011, Misbah2012, Tahiri2013, Aouane2014, Lazaro2014} as well as 3D \cite{Noguchi2005, Pivkin2008, McWhirter2011, Li2013, Fedosov2014,Sinha2015,Freund2014,Cordasco2017} simulations.
Our method allows for the full 3D capturing of moving bio-particles in micro-sized channels, which has so far been impossible (Fig.~\ref{fig2}).
We found two very different shapes of RBCs in the experiments: a croissant-like and a slipper-like shape. These observations are confirmed by 3D numerical simulations.

Our 3D imaging technique enables us to access the full 3D information at frame rates comparable to 2D approaches. In contrast to common $z$-scanning methods, we use a micro-fluidic channel which is tilted by a small angle with respect to the focal plane of a confocal microscope (Fig.~\ref{fig1}). This arrangement forces cells to pass the focal plane in an inclined manner and the acquisition of cross-sectional images (slices) of cells becomes possible without objective or stage motion. Due to their velocity, cells pass the field of view in the $x$-direction while the $z$-axis is automatically scanned in subsequent frames. In a post-processing step, the cell contour is then cropped from individual pictures and the volumetric representation assembled. Data acquisition is performed at a maximum rate of 600 frames per second (FPS) which facilitates to capture cells at velocities up to $1.5\,\text{mm/s}$. This mostly covers the physically relevant range of blood in the microvascular system. Limitations mainly arise from the maximum rotation speed of the Nipkow disks and the frame rate of the used camera. 

\begin{figure}[htb!]
	\centering
	\includegraphics{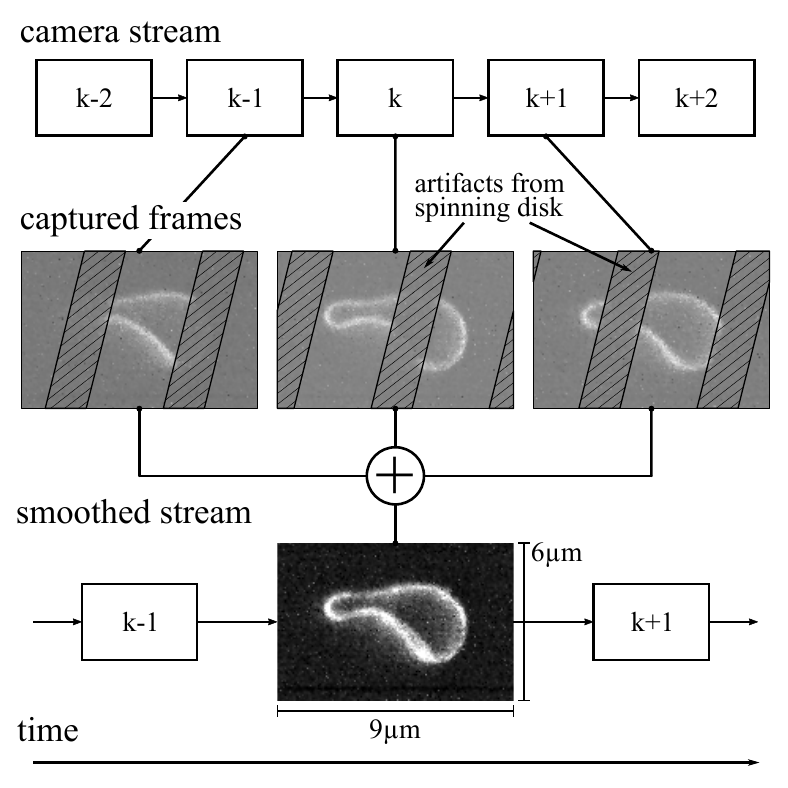}
	\caption{Post-processing of subsequent velocity corrected image frames (`camera stream') taken at discrete time-steps $k$. Artifacts from confocal imaging can exemplarily be seen for three subsequent cross-sectional images (`captured frames'). Areas (hatched) which are not excited by the light source and not seen by the camera remain dark and the observed cell reveals missing parts. Therefore, the superposition of three subsequent images is continuously built to get the full representation of the cell for each time-step (`smoothed stream'). As a side-effect, this procedure additionally improves the signal-to-noise ratio by a factor of $\approx 1.7$ and therefore enhances the image contrast.}
	\label{fig3}
\end{figure}

Special care is required when choosing the objective. Since channels are tilted, sufficient working distance of the objective is mandatory for mechanical reasons. Especially if channels with heights $>10\,\mu\text{m}$ are used, the tilt angle must be increased to fully capture the cross-section of the channel by the field of view. However, this can easily lead to mechanical collisions of the channel substrate and the objective housing. For our experiments, we found an inclination angle of approximately $3^{\circ}$ to be sufficient to fully take advantage of the available field of view ($225\,\mu\text{m}\times225\,\mu\text{m}$). Measurements are taken at a distance of $\approx 50\,\text{mm}$ away from the channel inlet. Our micro-fluidic channels are fabricated in a standard soft-lithography process and are made of PDMS based substrates attached to glass slides. The channel cross-section corresponds to $25\,\mu\text{m}\times10\,\mu\text{m}$ ($\text{width}\times\text{height}$). For our experiments, cells are stained with Cell-Mask\texttrademark{} Red which is excited at $647\,\text{nm}$. Using a standard protocol for staining, this dye very homogeneously attaches to the cell membrane and cross-sectional images clearly show the cell outlines (Fig.~\ref{fig3}). Achieving high selectivity between different slices, a $60\times$ oil immersion objective at high numerical aperture ($\text{NA}=1.2$) is used for confocal imaging. Aiming to characterize single cells, highly diluted RBC concentrations are used (approximately $10\,\mu$l blood suspended in $1\,\text{ml}$ PBS). Within the buffer solution, the viscosity contrast (ratio of inner and outer viscosity of cells) corresponds to $\lambda \approx 5$ which is in accordance with physiological conditions \cite{Cokelet1968}.

Confocal cross-sectional imaging is performed with a spinning micro-lens enhanced Nipkow disk. Rotating at $4000 \,\text{min}^{-1} \approx 67\,\text{s}^{-1}$ and equipped with multiple micro-lens sections, the disk allows for a nominal frame rate of $200\,\text{FPS}$. Since the camera exhibits an acquisition speed of $600\,\text{FPS}$, images are taken three times faster than the maximum rate specified for the disk. This results in noticeable dark areas from the micro-lens array that can be seen at individual image frames taken at a maximum of $600\,\text{FPS}$ (Fig.~\ref{fig3}) and mainly concerns the illumination as well as the fluorescence detection at certain image areas. For overcoming these disturbances, two obvious approaches exist. First, the camera could be synchronized with the rotating disk. Doing so, an overall sample rate of $200\,\text{FPS}$ would result which limits the observable velocity range of passing cells. Second, the camera can be decoupled from the disk, runs at maximum acquisition speed and records complete as well as fractional cell images (preferred method). Due to this oversampling, more information content on cells or fraction of cells can be gathered. However, for reconstructing a clean 3D representation of each cell, partial images must be supplemented by complete frames. This can be done by taking bundles of three subsequent cross-sectional slices which are continuously summarized to obtain the full representation of the cell membrane. Besides the effect that dark areas from confocal imaging are smoothed out, an improved signal-to-noise-ratio (factor $\approx 1.7$) results that enhances the image contrast.

Moreover, due to cell motion, the spatial shift of cell signals between subsequent frames must be determined for correct volumetric recovery. To access this shift, the cell velocity has to be measured. This is achieved by calculating the center of mass of individual cells for each time-step (frame) to obtain a position (pixel) vs.\ time diagram that is linearly fitted. Taking the magnification into account, the slope of this fit represents the cell velocity. Finally, a velocity corrected stack of cross-sectional cell images results which is then assembled to get the full 3D representation.

At equilibrium, shape assembly errors may only be due to statistical rotational or translational fluctuations caused by Brownian motion. Giving the reader a rough estimation on diffusion processes, spherical particles of similar volume serve as placeholder in the following calculation. Assuming a velocity of $350\,\mu\text{m/s}$, cells take $\tau=660\,\text{ms}$ to pass the field of view ($225\,\mu\text{m}$).
With a radius of $R=4\,\mu\text{m}$ and dynamic viscosity $\eta=1\,\text{mPas}$ of the medium, the root mean-square angular deviation during interaction corresponds to $\sqrt{\left\langle \Theta^{2} \right\rangle} = 0.057\,\text{rad}=3.25\,^\circ$. Moreover, the translational root mean square displacement amounts to $\sqrt{\left\langle x^{2}\right\rangle}=0.26\,\mu\text{m}$. Consequently, the rotational diffusion accounts for less than $1\%$ of a full rotation. Translational diffusion corresponds to $\approx6.5\%$ of the object size during its time of flight. Both diffusion processes may lead to slight displacements when the cross-sectional image slices are assembled. Fortunately, the main features of cell geometries are maintained and clear pictures evolve in practice. Thus, our measuring approach is in principle not effected by Brownian motion.

Fig.~\ref{fig2} (top row) shows typical 3D measurements for RBCs within a Poiseuille flow, comprising a slipper and a croissant shape as also observed in previous experiments \cite{Prado2015,Lanotte2014,Tomaiuolo2009,Abkarian2008,Gaehtgens1980} and simulations \cite{Fedosov2014,Secomb2007,Kaoui2009,Kaoui2011,Tahiri2013,Noguchi2005} (see the supplementary information for movies).
The slipper has a somewhat lower velocity ($\approx 330\,\mu\text{m}/\text{s}$) than the croissant ($\approx 370\,\mu\text{m}/\text{s}$), despite being driven by the same pressure drop, because it is slightly off-centered.
This is in agreement with previous 2D simulations \cite{Tahiri2013}.
Note that the croissant shape is similar to a parachute. But where the latter is perfectly rotationally symmetric around its axis, the croissant has only two symmetry planes. This is due to the rectangular form of the channel.

In addition to our experiments, we perform numerical simulations to confirm these shapes theoretically by using the periodic boundary integral method \cite{BubblePreprint2}. 
As it is based on the Stokes equation, the Reynolds number must be small in order to faithfully capture the dynamics.
For the present case it can be estimated to be $\mathrm{Re} = D_\mathrm{RBC} u \rho / \mu < 10^{-2}$, with the equilibrium diameter of the red blood cell $D_\mathrm{RBC} \approx 8\,\mu\text{m}$, the velocity $u \approx 370\,\mu\text{m}/\text{s}$, the density $\rho \approx 10^3 \,\text{kg}/\text{m}^3$ and the dynamic viscosity $\mu \approx 10^{-3}\, \text{kg}/(\text{s}\,\text{m})$.
The simulated channel has the same cross-section ($25\,\mu\text{m}\times10\,\mu\text{m}$) as in the experiments and a length of $60\,\mu\text{m}$. A single red blood cell is started in the equilibrium discocyte state and axis-aligned with the channel axis.
The viscosity at the inside is chosen to be 5 times higher than at the outside ($\lambda = 5$). For further details we refer the reader to our recent publications \cite{BubblePreprint2,Guckenberger2016,Guckenberger2017} and the supplementary information (SI). Depending on the starting position, either an off-centered and slower slipper or a centered and faster croissant can be observed (see the SI for movies). They compare favorably with the experimentally obtained forms as shown at the bottom of Fig.~\ref{fig2}. Note that the croissant shape is found to be metastable for the present parameters, i.e.\ it switches to a slipper shape after several seconds. This also suggests that the croissant as seen in the experiments might be of transient nature.
However, further studies are needed to confirm this observation. Moreover, the slipper was found to exhibit pronounced tank-treading in accordance with previous publications \cite{Tahiri2013,Fedosov2014}.

Summarizing, we developed a tomography based 3D imaging technique for moving bio-particles in micro-fluidic flow. Our approach will enable a large set of practically relevant application scenarios which have hitherto been inaccessible to confocal (or any other) 3D imaging technique. The key of our approach is a tilted channel that allows objects to traverse the field of view in an inclined manner. At sufficient frame rates, a number of cross-sectional images are taken which are then assembled to obtain the full 3D representation of bio-particles. This method enables us to characterize the shape of flowing cells, exemplified herein by using human RBCs. With 3D tomography we are able to confirm or refute theoretical and numerical predictions of cell shapes and can therefore provide more detailed information on cell-intrinsic parameters. In addition to our experimental technique, we performed numerical simulations of RBCs in the same environment as in the experiments. Employing the periodic boundary integral method, we found good agreement with the experimental results. Our approach could be utilized for diagnostics, e.g.\ to identify shape-related anomalies such as sickle-cell anemia. This makes micro-fluidic based 3D tomography a possible supplement or even replacement for flow cytometers in specific use-cases.

\section*{supplementary material}

See our supplementary material for a detailed description of the numerical simulation methods used as well as for videos of our results.

\begin{acknowledgments}

Funding from the Volkswagen Foundation and computing time granted by the Leibniz-Rechenzentrum on \mbox{SuperMUC} are gratefully acknowledged by A.~Guckenberger and S.~Gekle. S.~Quint, A.F.~Christ, S.~Himbert and C.~Wagner kindly acknowledge the support and funding of the `Deutsch-Franz\"osische-Hochschule'\ (DFH). Funding from the European Framework `Horizon 2020' under grant agreement number 675115 (RELEVANCE) is received by C.~Wagner and L.~Kaestner.
\end{acknowledgments}

\end{document}